# What is vacuum?

Peter Rowlands

*Department of Physics, University of Liverpool, Oliver Lodge Laboratory, Oxford Street, Liverpool, L69 7ZE, UK. e-mail p.rowlands@liverpool.ac.uk*

*Abstract*. Vacuum can be defined with exact mathematical precision as the state which remains when a fermion, with all its special characteristics, is created out of absolutely nothing. The definition leads to a special form of relativistic quantum mechanics, which only requires the construction of a creation operator. This form of quantum mechanics is especially powerful for analytic calculation, at the same time as explaining, from first principles, many aspects of the Standard Model of particle physics. In particular, the characteristics of the weak, strong and electric interactions can be derived from the structure of the creation operator itself.

**1 The nilpotent Dirac equation**

Vacuum is defined as the state of minimum (but seemingly nonzero) energy in quantum mechanics, it is also an active component in quantum field theory, while it is the main objective of such projected unifying theories as string theory to find the particular vacuum which makes their particle structures possible. For all its significance, however, vacuum is not a well-defined concept, and, though no one has found a way of eliminating it from quantum theories, the reason why nature requires it at all has never been made clear. However, it is possible to show that vacuum has an exact, mathematically precise and logically satisfying meaning, and that the discovery of that meaning is a very significant step in understanding the Standard Model of particle physics. Although this understanding requires a very particular formulation of relativistic quantum mechanics, it turns out that this is the most compact and powerful formulation available and the one that leads most readily into a quantum field representation and a rich field of interpretation in particle physics.

    Physics at its most fundamental level is entirely concerned with fermions and their interactions, gauge bosons being generated by such interactions. A quantum mechanical equation for the fermionic state might therefore be expected to give us a great deal of information about these interactions and related matters. The Dirac equation, in its usual form, certainly tells us how to handle the interactions in terms of calculation, but it tells us very little about their origins and distinctive characteristics. This may be partly due to the specific mathematical form of the equation, as normally used, and it may be that a greater depth of physical information may be revealed by

using a mathematical structure which is more transparent and easier to manipulate. Conventionally, we write the equation in the form

$$\left(\gamma^0 \frac{\partial}{\partial t} + \boldsymbol{\gamma}.\nabla + im\right)\psi = \left(\gamma^0 \frac{\partial}{\partial t} + \gamma^1 \frac{\partial}{\partial x} + \gamma^2 \frac{\partial}{\partial y} + \gamma^3 \frac{\partial}{\partial z} + im\right)\psi = 0 \quad (1)$$

where $\gamma^0$, $\gamma^1$, $\gamma^2$ and $\gamma^3$, are taken to be operators, which anticommute with each other, and with a fifth operator, $\gamma^5 = i\gamma^0\gamma^1\gamma^2\gamma^3$, and where

$$(\gamma^0)^2 = (\gamma^5)^2 = 1 \quad (\gamma^1)^2 = (\gamma^2)^2 = (\gamma^3)^2 = -1. \quad (2)$$

Usually, the $\gamma$ terms are taken to be 4 × 4 matrices, but this is an unnecessarily restrictive condition, and the only real requirements are that they are anticommuting operators with the multiplication properties defined in (2). In fact, since the $\gamma$ algebra is widely recognised as a Clifford algebra, it seems reasonable to use a more directly Cliffordian representation of the $\gamma$ operators, even though we could retain the $\gamma$ symbolism if required. The object here is not mathematical elegance but physical transparency, and so we construct an algebra which is closely related to the twistors of Penrose [1], and to Hestenes' multivariate vectors [2]. To create a system of five anticommuting operators in an associative Clifford algebra, we need at least two commuting 3-dimensional systems of units. The simplest choice which will reflect the physical properties of the terms involved appears to be a combination of quaternions and multivariate 4-vectors. The units of this algebra then become:

| | | | |
|---|---|---|---|
| *i j k* | quaternion units | **i j k** | multivariate vector units |
| 1 | scalar | *i* | pseudoscalar |

The quaternions (represented by bold italic symbols) obey the usual multiplication rules $\boldsymbol{i}^2 = \boldsymbol{j}^2 = \boldsymbol{k}^2 = \boldsymbol{ijk} = -1$, with a scalar term to complete the algebra, while the multivariate vectors (represented by bold symbols) are simply complexified quaternions, with the multiplication rules $\mathbf{i}^2 = \mathbf{j}^2 = \mathbf{k}^2 = -i\mathbf{ijk} = 1$, and a corresponding imaginary scalar (or pseudoscalar) to complete the algebra. In general, multivariate vectors **a** and **b** are distinguished from ordinary vectors by defining a full product:

$$\mathbf{ab} = \mathbf{a}.\mathbf{b} + i\,\mathbf{a} \times \mathbf{b}. \quad (3)$$

The units **i**, **j**, **k** are isomorphic to Pauli matrices, and Hestenes and others have shown that the additional cross-product term which appears if we make the $\nabla$ operator multivariate allows us to account for fermionic spin, even in the Schrödinger equation [2, 3]. In the Dirac equation, the same effect results from replacing $\boldsymbol{\gamma}.\nabla$ with $\boldsymbol{\gamma}\nabla$, where both $\boldsymbol{\gamma}$ and $\nabla$ are multivariate.

Just as the complete $\gamma$ algebra, with all possible permutations, has 64 units (including + and – terms), and forms a group of that order, of which the five $\gamma$ matrices are generators, so the eight units *i*, *j*, *k,* 1, **i**, **j**, **k**, *i* create a group of 64 possible combinations, which can be generated by five terms which are isomorphic to the $\gamma$ matrices. The mappings can be done in many different ways, but all are equivalent in total structure. We will find it convenient to define two, so that we can effect a transformation:

$$\gamma^0 = -i\mathbf{i} \quad \gamma^1 = \mathbf{i}\mathbf{k} \quad \gamma^2 = \mathbf{j}\mathbf{k} \quad \gamma^3 = \mathbf{k}\mathbf{k} \quad \gamma^5 = i\mathbf{j} \tag{4}$$

$$\gamma^0 = i\mathbf{k} \quad \gamma^1 = i\mathbf{i} \quad \gamma^2 = \mathbf{j}i \quad \gamma^3 = \mathbf{k}i \quad \gamma^5 = i\mathbf{j} \tag{5}$$

Substituting the first (4) into equation (1), we obtain

$$\left(-i\mathbf{i}\frac{\partial}{\partial t} + \mathbf{k}\mathbf{i}\frac{\partial}{\partial x} + \mathbf{k}\mathbf{j}\frac{\partial}{\partial y} + \mathbf{k}\mathbf{k}\frac{\partial}{\partial z} + im\right)\psi = 0. \tag{6}$$

A key move now is to multiply the equation from the left by $j$, at the same time altering the $\gamma$ representation to (5). After this transformation, the equation becomes:

$$\left(i\mathbf{k}\frac{\partial}{\partial t} + i\mathbf{i}\frac{\partial}{\partial x} + i\mathbf{j}\frac{\partial}{\partial y} + i\mathbf{k}\frac{\partial}{\partial z} + ijm\right)\psi = 0. \tag{7}$$

The remarkable thing about this equation is that it is fully symmetric. For the first time, all the $\gamma$ terms are incorporated into the equation on the same footing. Though the 3-dimensionality of the anticommutative operators $i, j, k$ and $\mathbf{i}, \mathbf{j}, \mathbf{k}$ makes their cyclic nature explicit, equation (7) does not depend on the algebraic representation. If we had been prepared to make the same change of representation which allows the transition from (6) to (7), we could have obtained the same result using the $\gamma$ notation.

$$-i\gamma^5\left(\gamma^0\frac{\partial}{\partial t} + \gamma^1\frac{\partial}{\partial x} + \gamma^2\frac{\partial}{\partial y} + \gamma^3\frac{\partial}{\partial z} + im\right)\psi = 0$$

$$\left(-i\gamma^5\gamma^0\frac{\partial}{\partial t} - i\gamma^5\gamma^1\frac{\partial}{\partial x} - i\gamma^5\gamma^2\frac{\partial}{\partial y} - i\gamma^5\gamma^3\frac{\partial}{\partial z} + \gamma^5 m\right)\psi = 0$$

$$\left(\gamma^0\frac{\partial}{\partial t} + \gamma^1\frac{\partial}{\partial x} + \gamma^2\frac{\partial}{\partial y} + \gamma^3\frac{\partial}{\partial z} + \gamma^5 m\right)\psi = 0. \tag{8}$$

Collecting the multivariate terms, for convenience, we can also write equations (7) and (8) in the form

$$\left(i\mathbf{k}\frac{\partial}{\partial t} + i\nabla + ijm\right)\psi = 0. \tag{9}$$

$$\left(\gamma^0\frac{\partial}{\partial t} + \boldsymbol{\gamma}\cdot\nabla + \gamma^5 m\right)\psi = 0 \tag{10}$$

Though these equations result only from algebraic transformations, they have a special *physical* significance. This becomes apparent when we apply a plane wave free particle solution to (9),

$$\psi = A e^{-i(Et - \mathbf{p}.\mathbf{r})}. \tag{11}$$

The result is an equation of the form

$$(kE + ii\mathbf{p} + ijm)\, A e^{-i(Et - \mathbf{p}.\mathbf{r})} = 0.$$

This equation is only valid if $A$ is either equal to $(kE + ii\mathbf{p} + ijm)$ or a nonquaternionic multiple of it, because $(kE + ii\mathbf{p} + ijm)$ is a *nilpotent*, a mathematical object that squares to zero, as in

$$(kE + ii\mathbf{p} + ijm)(kE + ii\mathbf{p} + ijm) = -E^2 + p^2 + m^2 = 0. \tag{12}$$

That is, we can write (11) in the form

$$\psi = (kE + ii\mathbf{p} + ijm) e^{-i(Et - \mathbf{p}.\mathbf{r})},$$

thus implying that a free particle wavefunction has a nilpotent amplitude. The same would be true if we had used equation (10) and the $\gamma$ operator notation, and it must have been true even before pre-multiplication of the operator by $j$ or $-i\gamma^5$. Nilpotency is a fundamental aspect of the free fermion state, but it is not just a mathematical condition; it also has an intrinsically *physical* meaning, and, as we will see, it applies as much to the bound or interacting, as well as to the free, fermion state.

## 2 The 4-component spinor

The conventional Dirac equation, of course, requires the wavefunction to be a *spinor*, with four components, structured as a column vector, representing the four combinations of particle and antiparticle, and spin up and spin down. Using $\pm E$ and a multivariate $\pm \mathbf{p}$ to represent these possibilities, we can easily see that the respective amplitudes of these states may be represented by

$$\begin{array}{c}
(kE + ii\mathbf{p} + ijm) \\
(kE - ii\mathbf{p} + ijm) \\
(-kE + ii\mathbf{p} + ijm) \\
(-kE - ii\mathbf{p} + ijm)
\end{array} \tag{13}$$

each multiplied by the appropriate phase factor. However, it is convenient at this point (for intrinsically physical reasons) to change the arbitrary sign convention which we have inherited from the conventional Dirac equation, and rewrite the column vector (13) in the form:

$$\begin{array}{c}
(ikE + i\mathbf{p} + jm) \\
(ikE - i\mathbf{p} + jm) \\
(-ikE + i\mathbf{p} + jm) \\
(-ikE - i\mathbf{p} + jm)
\end{array} \tag{14}$$

Applying the usual interpretations of the terms in the Dirac 4-spinor, we can identify these four states as representing, say,

$$(i\mathbf{k}E + i\mathbf{p} + jm) \qquad \text{fermion spin up}$$
$$(i\mathbf{k}E - i\mathbf{p} + jm) \qquad \text{fermion spin down}$$
$$(-i\mathbf{k}E + i\mathbf{p} + jm) \qquad \text{antifermion spin down}$$
$$(-i\mathbf{k}E - i\mathbf{p} + jm) \qquad \text{antifermion spin up}$$

The meanings associated with the signs of $E$ and $\mathbf{p}$, of course, are decided purely by convention, but once this is fixed, the spin state of the particle (or, more precisely, the helicity or handedness) is determined by the ratio of the signs of $E$ and $\mathbf{p}$. So $i\mathbf{p} / i\mathbf{k}E$ has the same helicity as $(-i\mathbf{p}) / (-i\mathbf{k}E)$, but the opposite helicity to $i\mathbf{p} / (-i\mathbf{k}E)$.

While conventional representations require different phase factors for positive and negative energy states, the current formalism allows us to use a *single* phase factor, if we structure the operator as a 4-component spinor, which we can represent as a row vector operating on the 4-component column vector forming the amplitude. Using the sign convention for amplitude as in (12), the corresponding row vector representing the differential operator would now be composed of the terms:

$$\left(-\mathbf{k}\frac{\partial}{\partial t} - i i \nabla + jm\right) \quad \left(-\mathbf{k}\frac{\partial}{\partial t} + i i \nabla + jm\right) \quad \left(\mathbf{k}\frac{\partial}{\partial t} - i i \nabla + jm\right) \quad \left(\mathbf{k}\frac{\partial}{\partial t} + i i \nabla + jm\right) \quad (15)$$

So an abbreviated form of the Dirac equation using a 4-component spinor operator and a 4-component spinor amplitude could be represented by:

$$\left(\mp \mathbf{k}\frac{\partial}{\partial t} \mp i i \nabla + jm\right)\left(\pm i\mathbf{k}E \pm i\mathbf{p} + jm\right)e^{-i(Et-\mathbf{p}\cdot\mathbf{r})} = 0. \qquad (16)$$

Of course, we can also use the symbols $E$ and $\mathbf{p}$ to represent the respective operators $i\partial / \partial t$ and $-i\nabla$, so equation (16) can also be written

$$\left(\pm i\mathbf{k}E \pm i\mathbf{p} + jm\right)\left(\pm i\mathbf{k}E \pm i\mathbf{p} + jm\right)e^{-i(Et-\mathbf{p}\cdot\mathbf{r})} = 0, \qquad (17)$$

where the terms in the first bracket represent operators and those in the second bracket eigenvalues. This suggests that we could derive the Dirac equation (16) simply by factorizing the Einstein energy-momentum relation, as in (12), and then applying a canonical quantization to the left-hand bracket.

There is one further approach which can reduce the amount of information needed to specify the quantum mechanics to just a two-term operator, eliminating both the mass and the phase factor. This is to use a discrete or anticommutative differentiation process, with a correspondingly discrete wavefunction. To define a discrete differentiation of function $F$, and also preserve the Leibniz chain rule, it is convenient to take:

$$\frac{\partial F}{\partial t} = [F, \mathcal{H}] = [F, E] \quad \text{and} \quad \frac{\partial F}{\partial X_i} = [F, P_i], \tag{18}$$

with $\mathcal{H} = E$ and $P_i$ representing energy and momentum operators [4]. (Here we assume that, with velocity operators not in evidence, we can use $\partial F / \partial t$ rather than $dF / dt$.) The mass term (which has only a passive role in quantum mechanics) disappears in the operator, though it has to be introduced in the amplitude. Suppose we define a nilpotent amplitude

$$\psi = i\mathbf{k}E + i\mathbf{i}P_1 + i\mathbf{j}P_2 + i\mathbf{k}P_3 + jm$$

and an operator

$$\mathcal{D} = i\mathbf{k}\frac{\partial}{\partial t} + i\mathbf{i}\frac{\partial}{\partial X_1} + i\mathbf{j}\frac{\partial}{\partial X_2} + i\mathbf{k}\frac{\partial}{\partial X_3},$$

with 
$$\frac{\partial \psi}{\partial t} = [\psi, \mathcal{H}] = [\psi, E] \quad \text{and} \quad \frac{\partial \psi}{\partial X_i} = [\psi, P_i],$$

With some straightforward algebraic manipulation, we find that

$$-\mathcal{D}\psi = i\psi(i\mathbf{k}E + i\mathbf{i}P_1 + i\mathbf{j}P_2 + i\mathbf{k}P_3 + jm)$$
$$+ i(i\mathbf{k}E + i\mathbf{i}P_1 + i\mathbf{j}P_2 + i\mathbf{k}P_3 + jm)\psi - 2i(E^2 - P_1^2 - P_2^2 - P_3^2 - m^2).$$

When $\psi$ is nilpotent, then

$$\mathcal{D}\psi = \left(i\mathbf{k}\frac{\partial}{\partial t} + i\nabla\right)\psi = 0.$$

If we generalise this to four states, with $\mathcal{D}$ and $\psi$ represented as 4-spinors, then

$$\mathcal{D}\psi = \left(\pm i\mathbf{k}\frac{\partial}{\partial t} \pm i\nabla\right)\left(\pm i\mathbf{k}E \pm i\mathbf{i}P_1 \pm i\mathbf{j}P_2 \pm i\mathbf{k}P_3 + jm\right) = 0 \tag{19}$$

becomes the equivalent of the nilpotent Dirac equation in this discrete calculus. It is significant that equations (18) did not employ the $i$ (or $i\hbar$) term usually required in canonical quantization to reach equation (19), though this could have been included, thus allowing a smooth transition between classical and quantum conditions. Again, this is only possible because of the nilpotency constraint.

**3 Vacuum**

The reduction to a single phase factor gives the formalism enormously increased calculating power, as finding this factor is the first objective of many calculations. Also, the correspondence in (14) and (15) between changes in the signs of $E$ and $t$ is a perfect illustration of the Feynman principle of particles having negative energy states also having reversed time direction. However, there is also a much more fundamental physical concept involved in the nilpotent structure of the wavefunction and the version of the Dirac equation represented in (17). Essentially, a particle with a nilpotent wavefunction, say $\psi_1$, will be automatically Pauli exclusive, because the combination state with an identical particle $\psi_1\psi_1$ will be zero. However, Pauli exclusion is not just true of free particles. In all cases where it has been observed, the fermions are interacting and subject to forces from other fermions. This is easily accommodated within the nilpotent formalism, as the operators $E$ and **p** need not represent just $i\partial / \partial t$ and $-i\nabla$, but can also incorporate field terms or covariant derivatives, so that $E$ could be, say, $i\partial / \partial t + e\phi +$ …, and **p** could be, say, $-i\nabla + e\mathbf{A} +$ … . The eigenvalues $E$ and **p** will then represent the more complicated expressions that will result from the presence of these terms. The phase factor will be changed from the $e^{-i(Et - \mathbf{p}\cdot\mathbf{r})}$ for the free particle, but the ultimate determining property of the system will be the need to maintain Pauli exclusion for all fermions, whether free or interacting.

In a formal sense, the reduction to a single phase factor and the extra constraint of nilpotency mean that much of the formal apparatus of relativistic quantum mechanics becomes redundant, in the sense that it need not be specified independently of the operator. If we write the operator in the form $(\pm ikE \pm i\mathbf{p} + jm)$, where $E$ and **p** are generic terms involving differentials and associated potentials, then the whole quantum mechanics of the system is completely specified. There is, strictly, no need for a wavefunction or even an equation. The operator alone uniquely determines the phase factor which is needed to create a nilpotent amplitude. Even the spinor representation is not strictly necessary as the first of the four terms, say ($ikE + i\mathbf{p} + jm$), uniquely specifies the remaining three by automatic sign variation, and it will often be convenient to specify the operator in this abbreviated form. We can suppose that it was the spinor structure of the original Dirac wavefunction which inhibited the development of a nilpotent formalism using $\gamma$ operators interpreted as matrices, even though, as equation (8) shows, this would have been technically possible, for, then matrices would have been required to exist *inside* a spinor already structured as a column vector, and then be acted on by a matrix differential operator.

If nilpotency is universal in fermion states, then we have an immediate understanding of the concept of vacuum, and also an immediate possibility of transformation from quantum mechanics to quantum field theory, without any formal process of second quantization. To understand vacuum, we simply imagine creating a fermion *ab initio*, that is, from absolutely nothing, with all the characteristics that we want to give it in terms of added potentials, interaction terms, etc. Vacuum is then simply the state that is left – everything other than the fermion. If, then, the wavefunction of the fermion is, say, $\psi_f$, the wavefunction of vacuum will be $\psi_v = -\psi_f$. The superposition will be the zero state we started from, $\psi_f + \psi_v = \psi_f - \psi_f = 0$, and, because the fermion is a nilpotent, the combination state

$$\psi_f\psi_v = -\psi_f\psi_f = -(\pm ikE \pm i\mathbf{p} + jm)(\pm ikE \pm i\mathbf{p} + jm)$$

will also be zero. Vacuum, in this understanding, becomes the 'hole' in the zero state produced by the creation of the fermion, or, from another point of view, the 'rest of the universe' that the fermion sees and interacts with. So, if we define a fermion with interacting field terms, then the 'rest of the universe' needs to be 'constructed' to make the existence of a fermion in that state possible.

Vacuum defined in this way requires a zero totality universe, a possibility that is now very seriously considered, especially in relation to a universe beginning *ab initio*. A zero condition for the entire universe is logically satisfying because it is necessarily *incapable* of further explanation. It is also a powerful route to understanding fundamental physical concepts because vacuum now becomes an active component of the theory. Here, it is important to realise that nilpotency is a statement of a *physical* principle, rather than a purely mathematical operation. Conventional relativistic quantum mechanics has been assumed to require idempotent, rather than nilpotent wavefunctions, i.e ones that square to themselves ($\psi\psi \to \psi$) rather than to zero, but exactly the same equation can be read as either idempotent or nilpotent, by a simple redistribution of a single algebraic unit between the sections of the equation defined as operator and wavefunction:

$$[(\mp \mathbf{k}\partial/\partial t \mp i\mathbf{i}\nabla + j m)\mathbf{j}] \, [\mathbf{j}(\pm i\mathbf{k}E \pm i\mathbf{p} + jm) \, e^{-i(Et - \mathbf{p}\cdot\mathbf{r})}] = 0.$$
$$\quad\quad\quad operator \quad\quad\quad\quad\quad idempotent \; wavefunction$$

$$[(\mp \mathbf{k}\partial/\partial t \mp i\mathbf{i}\nabla + j m)\mathbf{j}\mathbf{j}] \, [(\pm i\mathbf{k}E \pm i\mathbf{p} + jm) \, e^{-i(Et - \mathbf{p}\cdot\mathbf{r})}] = 0.$$
$$\quad\quad\quad operator \quad\quad\quad\quad\quad nilpotent \; wavefunction$$

Either interpretation is possible on a mathematical basis, and the idempotents will be shown later to have a special physical significance, but only the nilpotent reading of the equation gives us the rich structure of physical interpretation and powerfully simple methods of calculation that are described in this paper. The choice of formalism used in quantum mechanics is not a neutral one. Different mathematical structures reveal non-equivalent levels of physical information.

The nilpotent formalism reveals that a fermion 'constructs' its own vacuum, or the entire 'universe' in which it operates, and we can consider the vacuum to be 'delocalised' to the extent that the fermion is 'localised'. Clearly, no two fermions can have the same vacuum; the vacuum for one fermion cannot act as the vacuum for another. The 'local' can be defined as whatever happens inside the nilpotent structure ($\pm i\mathbf{k}E \pm i\mathbf{p} + jm$), and the 'nonlocal' as whatever happens outside it. A 'one fermion' theory of the universe, as originally proposed by Wheeler, and reported subsequently by Feynman [5], is therefore a serious possibility. However, a single fermion cannot be considered isolated. It must be interacting. In effect, it must construct a 'space', so that its vacuum is not localised on itself. If a fermion is point-like, its vacuum must be dispersed. In this sense, a single (noninteracting) fermion cannot exist. It can only be defined if we also define its vacuum.

Since Pauli exclusion is automatic with nilpotent wavefunctions, it is important to note that nilpotent wavefunctions or amplitudes are also Pauli exclusive in the conventional sense by being automatically antisymmetric, with nonzero

$$\psi_1 \psi_2 - \psi_2 \psi_1 = -(\psi_2 \psi_1 - \psi_1 \psi_2)$$

since

$$(\pm ikE_1 \pm i\mathbf{p}_1 + jm_1)(\pm ikE_2 \pm i\mathbf{p}_2 + jm_2)$$
$$- (\pm ikE_2 \pm i\mathbf{p}_2 + jm_2)(\pm ikE_1 \pm i\mathbf{p}_1 + jm_1)$$
$$= 4\mathbf{p}_1\mathbf{p}_2 - 4\mathbf{p}_2\mathbf{p}_1 = 8 i \, \mathbf{p}_1 \times \mathbf{p}_2. \tag{20}$$

The result, however, is quite remarkable, as it implies that, instantaneously, any nilpotent wavefunction must have a **p** vector in spin space (a kind of spin 'phase') at a different orientation to any other. The wavefunctions of all nilpotent fermions might then instantaneously correlate because the planes of their **p** vector directions must all intersect, and the intersections actually create the *meaning* of Euclidean space, with an intrinsic spherical symmetry generated by the fermions themselves.

**4 Spin**

The nilpotent operator ($ikE$ + $i\mathbf{p}$ + $jm$) immediately presents us with a Hamiltonian specified as $\mathcal{H}$ = ($i\mathbf{p}$ + $jm$) (or $-ik$ times this), and using this Hamiltonian, we can proceed to derive fermionic spin as a result of the multivariate nature of the **p** component. If we mathematically define a quantity $\boldsymbol{\sigma}$ = −**1** (a pseudovector of magnitude −1), then

$$[\boldsymbol{\sigma}, \mathcal{H}] = [-\mathbf{1}, i\,(\mathbf{i}p_1 + \mathbf{j}p_2 + \mathbf{k}p_3) + jm] = [-\mathbf{1}, i(\mathbf{i}p_1 + \mathbf{j}p_2 + \mathbf{k}p_3)]$$
$$= -2i\,(\mathbf{ij}p_2 + \mathbf{ik}p_3 + \mathbf{ji}p_1 + \mathbf{j}\,p_3 + \mathbf{ki}p_1 + \mathbf{kj}p_2)$$
$$= -2ii\,(\mathbf{k}(p_2 - p_1) + \mathbf{j}(p_1 - p_3) + \mathbf{i}(p_3 - p_2))$$
$$= -2ii\,\mathbf{1} \times \mathbf{p}.$$

If **L** is the orbital angular momentum **r** × **p**, then

$$[\mathbf{L}, \mathcal{H}] = [\mathbf{r} \times \mathbf{p}, i\,(\mathbf{i}p_1 + \mathbf{j}p_2 + \mathbf{k}p_3) + km]$$
$$= [\mathbf{r} \times \mathbf{p}, i\,(\mathbf{i}p_1 + \mathbf{j}p_2 + \mathbf{k}p_3)]$$
$$= i\,[\mathbf{r}, (\mathbf{i}p_1 + \mathbf{j}p_2 + \mathbf{k}p_3)] \times \mathbf{p}$$

But $\quad [\mathbf{r}, (\mathbf{i}p_1 + \mathbf{j}p_2 + \mathbf{k}p_3)]y = i\mathbf{1}\,y$.

Hence $\quad [\mathbf{L}, \mathcal{H}] = ii\,\mathbf{1} \times \mathbf{p},$

and $\mathbf{L} + \boldsymbol{\sigma} / 2$ is a constant of the motion, because

$$[\mathbf{L} + \boldsymbol{\sigma} / 2, \mathcal{H}] = 0.$$

The spin ½ term characteristic of fermionic states, which emerges from this formalism, has often been considered a rather strange property in seemingly requiring a fermion to undergo a $4\pi$, rather than $2\pi$, rotation to return to its starting point. However, if we regard a fermion as only being created simultaneously with its mirror image vacuum state, then we can regard the spin ½ term as an indication that taking the fermion alone only gives us half of the knowledge we require to specify the system. The spin of fermion plus vacuum is, of course, single-valued (0).

Helicity ($\sigma.\mathbf{p}$) is another constant of the motion because

$$[\sigma.\mathbf{p}, \mathcal{H}] = [-p, i\, (\mathbf{i}p_1 + \mathbf{j}p_2 + \mathbf{k}p_3) + ijm] = 0$$

It is significant that, for a multivariate $\mathbf{p}$,

$$\mathbf{pp} = (\sigma.\mathbf{p})(\sigma.\mathbf{p}) = pp = p^2$$

which means that we can also use $\sigma.\mathbf{p}$ ($\sigma p$) for $\mathbf{p}$ (or $\sigma.\nabla$ ($\sigma\nabla$) for $\nabla$) in the nilpotent operator. For a hypothetical fermion / antifermion with zero mass, we are reduced to two options or distinguishable states:

$$(ikE + i\,\sigma.\mathbf{p} + jm) \rightarrow (ikE - ip)$$
$$(-ikE + i\,\sigma.\mathbf{p} + jm) \rightarrow (-ikE - ip) \qquad (21)$$

Each of these terms is associated with a single sign of helicity, ($ikE + ip$) and ($-ikE + ip$) being excluded, if we choose the same sign conventions for $\mathbf{p}$. Because we were required to choose $\sigma = -\mathbf{1}$ in deriving spin for states with positive energy, the allowed spin direction for these states must be antiparallel, and so require left-handed helicity, while the helicity of the negative energy states becomes right-handed. Numerically, $\pm E = p$, so we can express the allowed states as $\pm E(\mathbf{k} - i\mathbf{i})$. Multiplication from the left by the projection operator $(1 - ij) / 2 \equiv (1 - \gamma^5) / 2$ then leaves the allowed states unchanged while zeroing the excluded ones.

Because of the way spin emerges from the specifically multivariate aspect of the operator $\mathbf{p}$, through the additional cross product term with its imaginary coefficient or pseudovector, as in (3), it is important to distinguish those cases where the space variables are multivariate from those where they are not, as, for example, when polar coordinates are used. In these cases, an *explicit* spin (or total angular momentum) term must be included. Dirac, for example, has given a prescription for translating his equation into polar form [6], and it is easy to see that, in this formalism, it will require the momentum operator to acquire an additional (imaginary) spin (or total angular momentum) term, so that, just as in the conventional Dirac approach:

$$\nabla \rightarrow \left(\frac{\partial}{\partial r} + \frac{1}{r}\right) \pm i\,\frac{j + \frac{1}{2}}{r}. \qquad (22)$$

Yet another significant aspect of spin emerges when we write a nilpotent Hamiltonian in the form

$$\mathcal{H} = -ijc\sigma.\mathbf{p} - iiimc^2 = -ijc\mathbf{1p} - iiimc^2 = \alpha c\mathbf{p} - iiimc^2,$$

using the original Dirac sign convention, and with the constant *c* now specifically included (as $\hbar$ will be later). Since we have four separate spin states in the system, $\alpha = -i\mathbf{j}\mathbf{1}$ may be taken as a dynamical variable, and $c\alpha = -i\mathbf{j}\mathbf{1}c$ defined as a velocity operator, which, for a free particle, becomes:

$$\mathbf{v} = \dot{\mathbf{r}} = \frac{d\mathbf{r}}{dt} = \frac{1}{i\hbar} [\mathbf{r}, \mathcal{H}] = -i\mathbf{j}\mathbf{1}c = c\alpha .$$

Here, again, we have used the discrete calculus of equations (18), though the use of an explicit velocity operator now requires $dF / dt$ to be distinguished from $\partial F / \partial t$. The equation of motion for this operator then becomes:

$$\frac{d\alpha}{dt} = \frac{1}{i\hbar} [\alpha, \mathcal{H}] = \frac{2}{i\hbar} (c\mathbf{p} - \mathcal{H}\alpha).$$

This is, of course, a standard result, and the solution, giving the equation of motion for the fermion, was first obtained by Schrödinger [7]:

$$\mathbf{r}(t) = \mathbf{r}(0) + \frac{c^2 \mathbf{p}}{\mathcal{H}} t + \frac{\hbar c}{2i\mathcal{H}} [\alpha(0) - c\mathcal{H}^{-1}\mathbf{p}](\exp (2i\mathcal{H}t / h) - 1).$$

Only the third term has no classical analogue, seemingly predicting a violent oscillatory motion or high-frequency vibration (*zitterbewegung*) of the particle at frequency $\approx 2mc^2 / \hbar$, and amplitude $\hbar / 2mc$, which is related to the Compton wavelength for the particle and directly determined by the particle's rest mass. Since the term is derived from a velocity operator, defined as $c\alpha = -i\mathbf{j}\mathbf{1}$, the *zitterbewegung* has always been interpreted as a switching between the fermion's four spin states. It is certainly a vacuum effect. Its representation of the continual re-enactment of fermion creation is the most definite statement of the vacuum's existence. In relation to the nilpotent formalism, it seems to provide a physical mechanism for accommodating the instantaneous spin phase required by equation (20). This is, additionally, possible if we imagine an alternative representation of nilpotency as representing a unique direction on a set of axes defined by the values of *E*, **p** and *m*. In such a representation, half of the possibilities on one axis (those with –*m*) would be eliminated automatically (as being in the same direction as those with *m*), as would all those with zero *m* (since the directions would all be along the line *E* = *p*); such hypothetical massless particles would be impossible, in addition, for fermions and antifermions with the same helicity, as *E*, *p* has the same direction as –*E*, –*p*.

**5 Quantum mechanics and the quantum field**

The nilpotent formalism is one in which multiple physical meanings are encoded within the symbols. The nilpotent condition itself,

$$(\pm ikE \pm i\mathbf{p} + jm)(\pm ikE \pm i\mathbf{p} + jm) \to 0,$$

can be interpreted in many different ways, depending on the specific meaning of the symbols in the brackets, for example:

| | |
|---|---|
| classical variables | special relativity |
| operator × operator | Klein-Gordon equation |
| operator × wavefunction | Dirac equation |
| wavefunction × wavefunction | Pauli exclusion |
| wavefunction × (–) wavefunction | fermion and vacuum |

All the meanings are, of course, connected, and they also seem to encode other important aspects of physics. The fermion and vacuum connection, for example, which implies that a fermion can only be described with respect to the rest of the universe, implies a significant amount of thermodynamics, for it requires conservation of energy at all times (the first law), while denying that the fermion can ever be part of a closed system (the second law). The fermion necessarily defines an open system, and the thermodynamics of any observed system will necessarily be of a nonequilibrium nature.

The most significant aspect of the nilpotent formalism, however, is that it is already a full quantum field theory in which the operators act on the entire quantum field, without needing any formal process of second quantization. A nilpotent operator, once defined, acts as a creation operator acting on vacuum to create the fermion, together with all the interactions in which it is involved. The point of transformation from quantum mechanics to quantum field theory is the point at which we choose to privilege the operator rather than the equation, and at which we assume that Pauli exclusion applies to all fermionic states, whether free or bound, and regardless of the number of interactions to which they are subject. No additional mathematical formalism is necessary. As we have seen, once we have taken this simple, but profound step, we no longer need an equation at all. We simply define a fermion creation operator in differential form and imagine creating it from nothing. The phase factor is simply an expression of all the possible variations in space and time which are encoded in the creation operator. This is uniquely defined once the operator is specified. A fermion is thus specified as a set of space and time variations. The mass term, as we see from equation (19), is purely passive, and is convenient, rather than necessary information.

The special advantage of the formalism is that it contains all the information required of a quantum field theory, while retaining the simpler structures of quantum mechanics in the conventional sense, and it is, of course, possible to use it to do quantum mechanics. Here, it is most convenient to define a probability density for a nilpotent wavefunction $(\pm ikE \pm i\mathbf{p} + jm)$ by multiplication with its *complex quaternion conjugate* $(\pm ikE \mp i\mathbf{p} - jm)$ (the extra 'quaternion' resulting from the fact that the nilpotent wavefunction differs from a conventional one through premultiplication by a quaternion operator). So the unit probability density can be defined by

$$\frac{(\pm ikE \pm i\mathbf{p} + jm)}{\sqrt{2E}} \frac{(\pm ikE \mp i\mathbf{p} - jm)}{\sqrt{2E}} = 1,$$

the $1/\sqrt{2E}$ being a normalizing factor. If such factors are automatically assumed to apply in calculations, we can also define $(\pm ikE \mp i\mathbf{p} - jm)$ as the 'reciprocal' of $(\pm ikE \pm i\mathbf{p} + jm)$.

Individual calculations are also possible, in many cases using many fewer steps than by any conventional process, and sometimes also providing extra physical information. (The ease of calculation is clearly related to the fact that *dual* information, concerning both fermion and vacuum, is available.) We may, for example, consider using (22) to write down a non-time-varying nilpotent operator in polar coordinates:

$$(ikE - i i \nabla + jm) \rightarrow \left( ikE - i i \left( \frac{\partial}{\partial r} + \frac{1}{r} \pm i \frac{j + \frac{1}{2}}{r} \right) + jm \right). \tag{23}$$

Now, if the use of polar coordinates can be considered to represent spherical symmetry with respect to a point source, then (23) has no nilpotent solutions unless the *E* term also contains an expression proportional to $1/r$. In other words, simply defining a point source forces us to assume that a Coulomb interaction component is necessary for any nilpotent fermion defined with respect to it. In fact, all known forces have such components, together with an associated *U*(1) symmetry. For the gravitational and electric forces, it is the main or complete description; for the strong force it is the one-gluon exchange; for the weak field it is the hypercharge and the $B^0$ gauge field. Its effect is connected purely with scale or magnitude and we can associate it with the coupling constant. The fact is, of course, well known, and the inverse square relation was connected with the 3-dimensionality of space by Kant as early as the eighteenth century, but it is not a deductive consequence of any other existing physical theory.

If we now write the nilpotent operator in (24) with the required Coulomb term, we will find that it can be solved, using the known procedures, but eliminating many unnecessary ones, in only six lines of calculation. We begin with:

$$\left( \pm ik \left( E - \frac{A}{r} \right) \mp i i \left( \frac{\partial}{\partial r} + \frac{1}{r} \pm i \frac{j + \frac{1}{2}}{r} \right) + jm \right). \tag{24}$$

with a main requirement to find the phase factor $\phi$ which will make the amplitude nilpotent. So, we try the standard solution:

$$\phi = e^{-ar} r^{\gamma} \sum_{\nu=0} a_{\nu} r^{\nu}.$$

We then apply the operator in (24) to $\phi$, and square the result to 0 to obtain:

$$4\left(E-\frac{A}{r}\right)^2 = -2\left(-a+\frac{\gamma}{r}+\frac{v}{r}+\ldots\frac{1}{r}+i\frac{j+\frac{1}{2}}{r}\right)^2 - 2\left(-a+\frac{\gamma}{r}+\frac{v}{r}+\ldots\frac{1}{r}-i\frac{j+\frac{1}{2}}{r}\right)^2 + 4m^2.$$

Equating constant terms leads to
$$a = \sqrt{m^2 - E^2}. \tag{25}$$

Equating terms in $1/r^2$, following standard procedure, with $v = 0$, we obtain:

$$\left(\frac{A}{r}\right)^2 = -\left(\frac{\gamma+1}{r}\right)^2 + \left(\frac{j+\frac{1}{2}}{r}\right)^2. \tag{26}$$

Assuming the power series terminates at $n'$, following another standard procedure, and equating coefficients of $1/r$ for $v = n'$,

$$2EA = -2\sqrt{m^2 - E^2}\,(\gamma + 1 + n'), \tag{27}$$

the terms in $(j + \frac{1}{2})$ cancelling over the summation of the four multiplications, with two positive and two negative. Algebraic rearrangement of (25)-(27) then yields

$$\frac{E}{m} = \frac{1}{\sqrt{1+\dfrac{A^2}{(\gamma+1+n')^2}}} = \frac{1}{\sqrt{1+\dfrac{A^2}{\left(\sqrt{(j+\frac{1}{2})^2 - A^2} + n'\right)^2}}},$$

which, with $A = Ze^2$, becomes the hyperfine or fine structure formula for a one-electron nuclear atom or ion.

## 6 Bosons

The three quaternion units in the nilpotent operator also have multiple, but connected, meanings. One of these is as operators for fundamental symmetry transformations, by pre- and post-multiplication of the nilpotent operator.

| | | |
|---|---|---|
| Parity | $P$ | $\mathbf{i}\,(ik E + i\mathbf{p} + j m)\,\mathbf{i} = (ik E - i\mathbf{p} + j m)$ |
| Time reversal | $T$ | $\mathbf{k}\,(ik E + i\mathbf{p} + j m)\,\mathbf{k} = (-ik E + i\mathbf{p} + j m)$ |
| Charge conjugation | $C$ | $-\mathbf{j}\,(ik E + i\mathbf{p} + j m)\,\mathbf{j} = (-ik E - i\mathbf{p} + j m)$     (28) |

$TCP \equiv CPT \equiv$ identity is an automatic consequence from these conditions, because

$$\mathbf{k}\,(-\mathbf{j}\,(\mathbf{i}\,(ikE + i\mathbf{p} + jm)\,\mathbf{i})\,\mathbf{j})\,\mathbf{k} = -\mathbf{kji}\,(ikE + i\mathbf{p} + jm)\,\mathbf{ijk} = (ikE + i\mathbf{p} + jm),$$

as also are $CP \equiv T$, $PT \equiv C$, and $CT \equiv P$.

It is significant here that charge conjugation is effectively defined in terms of parity and time reversal, rather than being an independent operation. This is because only space and time are active elements, the variation in space and time being the coded information that solely determines the phase factor and the entire nature of the fermion state, and the mass term (which connects with the charge conjugation transformation) being a passive element, which can even be excluded from the operator without loss of information. It is relevant here that the construction of a nilpotent amplitude effectively requires the loss of a sign degree of freedom in one component, $E$, **p** or $m$, and that the passivity of mass makes it the term to which this will apply.

The representations of *P*, *T* and *C* symmetry transformations in (28) indicate something about the nature of the terms in the nilpotent 4-spinor, other than the lead term which determines the nature of the 'real' particle state. They are effectively, the *P*-, *T*- and *C*-transformed versions of this state, the states into which it could transform without changing the magnitude of its energy or momentum. We could perceive them as vacuum 'reflections' of the real particle state, and we will show in section 8 how they arise from vacuum operations that can be mathematically defined. Now, although Pauli exclusion prevents a fermion from forming a combination state with itself, we can imagine it forming a combination state with each of these vacuum 'reflections', and, if the 'reflection' exists or materialises as a 'real' state, then the combined state can form one of the three classes of bosons or boson-like objects.

A spin 1 boson can be imagined as being formed from a combination of fermion and antifermion with the same spins but opposite helicities. We take, for example, the product of a row vector fermion and a column vector antifermion, both written as columns for convenience:

$$
\begin{array}{ll}
(ikE + i\,\mathbf{p} + j\,m) & (-ikE + i\,\mathbf{p} + j\,m) \\
(ikE - i\,\mathbf{p} + j\,m) & (-ikE - i\,\mathbf{p} + j\,m) \\
(-ikE + i\,\mathbf{p} + j\,m) & (ikE + i\,\mathbf{p} + j\,m) \\
(-ikE - i\,\mathbf{p} + j\,m) & (ikE - i\,\mathbf{p} + j\,m).
\end{array} \qquad (29)
$$

The antifermion structure simply reverses the signs of $E$ throughout. The phase factor of both components is, according to our original construction of the nilpotent formalism, the same, dependent on the values of $E$ and **p** but not on their signs. The product is clearly a nonzero scalar (as the sign variations ensure cancellation of all the terms with quaternion coefficients), and so fulfils the condition for a boson wavefunction. Clearly, the same result will be obtained if the spin 1 boson is massless (as is the case with such gauge bosons as photons and gluons). Then we have:

$$
\begin{array}{ll}
(ikE + i\,\mathbf{p}) & (-ikE + i\,\mathbf{p}) \\
(ikE - i\,\mathbf{p}) & (-ikE - i\,\mathbf{p}) \\
(-ikE + i\,\mathbf{p}) & (ikE + i\,\mathbf{p}) \\
(-ikE - i\,\mathbf{p}) & (ikE - i\,\mathbf{p}).
\end{array} \qquad (30)
$$

The spin 0 boson is obtained by reversing the **p** signs in either fermion or antifermion, so that the components have the opposite spins but the same helicities:

$$
\begin{array}{ll}
(ikE + i\mathbf{p} + jm) & (-ikE - i\mathbf{p} + jm) \\
(ikE - i\mathbf{p} + jm) & (-ikE + i\mathbf{p} + jm) \\
(-ikE + i\mathbf{p} + jm) & (ikE - i\mathbf{p} + jm) \\
(-ikE - i\mathbf{p} + jm) & (ikE + i\mathbf{p} + jm).
\end{array}
\qquad (31)
$$

Again this gives a nonscalar scalar value, as required. However, this time, the mass cannot be reduced to zero, as nilpotency rules zero the product as well.

$$
\begin{array}{ll}
(ikE + i\mathbf{p}) & (-ikE - i\mathbf{p}) \\
(ikE - i\mathbf{p}) & (-ikE + i\mathbf{p}) \\
(-ikE + i\mathbf{p}) & (ikE - i\mathbf{p}) \\
(-ikE - i\mathbf{p}) & (ikE + i\mathbf{p}).
\end{array}
\qquad (32)
$$

Effectively, then, a spin 0 boson, defined by this process, cannot be massless. Hence, Goldstone bosons cannot exist, and the Higgs boson must have a mass. This mass is, additionally, as will become evident, a measure of the degree of right-handedness in the fermion component and left-handedness in the antifermion component.

A third possibility is a boson-like state formed by combining two fermions with opposite spins and opposite helicities:

$$
\begin{array}{ll}
(ikE + i\mathbf{p} + jm) & (ikE - i\mathbf{p} + jm) \\
(ikE - i\mathbf{p} + jm) & (ikE + i\mathbf{p} + jm) \\
(-ikE + i\mathbf{p} + jm) & (-ikE - i\mathbf{p} + jm) \\
(-ikE - i\mathbf{p} + jm) & (-ikE + i\mathbf{p} + jm).
\end{array}
\qquad (33)
$$

States of this nature can be imagined to occur in Cooper pairing in superconductors, in He$^4$ and Bose-Einstein condensates, in spin 0 nuclei, in the Jahn-Teller effect, the Aharonov-Bohm effect, the quantum Hall effect (where the second 'fermion' is a magnetic flux line), and, in general, in states where there is a nonzero Berry phase to make fermions become single-valued in terms of spin. In general, these will be spin 0 states, but they could become spin 1 states if, as is the case with He$^3$, the two components move with respect to each other, presumably in some kind of harmonic oscillator fashion, meaning that they could have the same spin states but opposite helicities. If they are spin 0, they can also have zero effective mass, as in Cooper pairing.

Now, the weak interaction can be considered as one in which fermions and antifermions are annihilated while bosons are created, or bosons are annihilated while fermions and antifermions are created, and, more generally, as one in which both processes (or equivalent) occur. As a creator and annihilator of states, it has the action of a harmonic oscillator. One of the fundamental differences between fermions and bosons is that fermions are sources for weak interactions, while bosons are not. Bosons, considered as created at fermion-antifermion vertices, are the products of weak interactions. Even in examples such as electron-positron collisions, where the predominant interaction is electric at low energies, there is an amplitude for a weak interaction. If we consider (29)-(33) as defining the vertices for boson production via the weak interaction, then it appears from (32) and from (21) that the pure weak interaction requires left-handed fermions and right-handed antifermions. In other words, it requires both a charge-conjugation violation and a simultaneous parity or time-reversal violation.

We can see in principle how this leads to mass generation by some process at least resembling the Higgs mechanism. Suppose we imagine a fermionic vacuum state with zero mass, say ($ikE + i\mathbf{p}$). An ideal vacuum would maintain exact and absolute *C*, *P* and *T* symmetries. Under *C* transformation, ($ikE + i\mathbf{p}$) would become ($-ikE - i\mathbf{p}$), with which it would be indistinguishable under normalization. No bosonic state would be required for the transformation, because the states would be identical. If, however, the vacuum state is degenerate in some way under charge conjugation (as supposed in the weak interaction), then ($ikE + i\mathbf{p}$) will be transformable into a state which can be distinguished from it, and the bosonic state ($ikE + i\mathbf{p}$) ($-ikE - i\mathbf{p}$) will necessarily exist. However, this can only be true if the state has nonzero mass and becomes the spin 0 'Higgs boson' ($ikE + i\mathbf{p} + jm$) ($-ikE - i\mathbf{p} + jm$). The mechanism, which produces this state, and removes the masslessness of the boson, requires the fixing of a gauge for the weak interaction (a 'filled' weak vacuum), which manifests itself in the massive intermediate bosons, *W* and *Z*.

The structures of bosons and the consideration of spin in section 4 suggest that mass and helicity are closely related. If the degree of left-handed helicity is determined by the ratio ($\pm$) $i\mathbf{p}$ / ($\pm$) $ikE$, then the addition of a mass term will change this ratio. Similarly, a change in the helicity ratio will also affect the mass. If the weak interaction is only responsive to left-handed helicity states in fermions, then right-handed states will be intrinsically passive, so having no other function except to generate mass. The presence of two helicity states will be a signature of the presence of mass. The *SU*(2) of weak isospin, which, in effect, expresses the invariability of the weak interaction to the addition of an opposite degree of helicity (due to the presence of, say, mass or electric charge) is thus related indirectly to the *SU*(2) of spin, which is a simple description of the existence of two helicity states. It is significant that the *zitterbewegung* frequency, which is a measure of the switching of helicity states, depends only on the fermion's mass. Mass is in some sense created by it, or is in some sense an expression of it. The restructuring of space and time variation or energy and momentum, via the phase factor, during an interaction, leads to a creation or annihilation of mass, which manifests itself in the restructuring of the *zitterbewegung*.

The coupling of a massless fermion, say ($ikE_1 + i\mathbf{p}_1$), to a Higgs boson, say ($ikE + i\mathbf{p} + jm$) ($-ikE - i\mathbf{p} + jm$), to produce a massive fermion, say ($ikE_2 + i\mathbf{p}_2 + jm_2$), can be imagined as occurring at a vertex between the created fermion ($ikE_2 + i\mathbf{p}_2 + jm_2$) and the antistate ($-ikE_1 - i\mathbf{p}_1$), to the annihilated massless fermion, with subsequent equalization of energy and momentum states. If we imagine a vertex involving a fermion superposing ($ikE + i\mathbf{p} + jm$) and ($ikE - i\mathbf{p} + jm$) with an antifermion superposing ($-ikE + i\mathbf{p} + jm$) and ($-ikE - i\mathbf{p} + jm$), then there will be a minimum of two spin 1 combinations and two spin 0 combinations, meaning that the vertex will be massive (with Higgs coupling) and carry a non-weak (i.e. electric) charge. So, a process such as a weak isospin transition, which, to use a very basic model, converts something like ($ikE_1 + i\mathbf{p}_1 + jm_1$) (representing isospin up) to something like $\alpha_1$ ($ikE_2 + i\mathbf{p}_2 + jm_2$) + $\alpha_1$ ($ikE_2 - i\mathbf{p}_2 + jm_2$) (representing isospin down), requires an additional Higgs boson vertex (spin 0) to accommodate the right-handed part of the isospin down state, when the left-handed part interacts weakly. This is, of course, what we mean when we say that the *W* and *Z* bosons have mass. The mass balance is done through separate vertices involving the Higgs boson.

**7 Baryons and gluons**

No fundamental explanation for baryon structure or the strong interaction has been previously proposed, but the nilpotent formalism suggests a mathematical representation of baryon structure which has exactly the required group characteristics. Here, we make the vector properties of **p** explicit so that we can write down a fermionic wavefunction with a 3-component structure. Clearly we cannot combine three components in the form:

$$(ikE \pm i\,\mathbf{p} + j\,m)\,(ikE \pm i\,\mathbf{p} + j\,m)\,(ikE \pm i\,\mathbf{p} + j\,m)$$

as this will immediately zero itself, but we can imagine one in which the vector nature of **p** plays an explicit role

$$(ikE \pm i\,\mathbf{i}p_x + j\,m)\,(ikE \pm i\,\mathbf{j}p_y + j\,m)\,(ikE \pm i\,\mathbf{k}p_z + j\,m)$$

and observe that it has nilpotent solutions when $\mathbf{p} = \pm i\,\mathbf{i}p_x$, $\mathbf{p} = \pm i\,\mathbf{j}p_y$, or $\mathbf{p} = \pm i\,\mathbf{k}p_z$, that is, when the momentum is directed entirely along the *x*, *y*, or *z* axes, in either direction, however defined. In principle, the complete wavefunction will contain the same information as if there were precisely six allowed independent phases, all existing simultaneously and subject to continual transitions at a constant rate. These six phases, which must be nonlocally gauge invariant, may be represented by:

$$
\begin{array}{llll}
(ikE + i\,\mathbf{i}p_x + j\,m) & (ikE + \ldots + j\,m) & (ikE + \ldots + j\,m) & +RGB \\
(ikE - i\,\mathbf{i}p_x + j\,m) & (ikE - \ldots + j\,m) & (ikE - \ldots + j\,m) & -RBG \\
(ikE + \ldots + j\,m) & (ikE + i\,\mathbf{j}p_y + j\,m) & (ikE + \ldots + j\,m) & +BRG \\
(ikE - \ldots + j\,m) & (ikE - i\,\mathbf{j}p_y + j\,m) & (ikE - \ldots + j\,m) & -GRB \\
(ikE + \ldots + j\,m) & (ikE + \ldots + j\,m) & (ikE + i\,\mathbf{k}p_z + j\,m) & +GBR \\
(ikE - \ldots + j\,m) & (ikE - \ldots + j\,m) & (ikE - i\,\mathbf{k}p_z + j\,m) & -BGR \quad (34)
\end{array}
$$

Using an appropriate normalization, these reduce to

$$(ikE + i\,\mathbf{i}p_x + \mathbf{j}\,m) \qquad +RGB$$
$$(ikE - i\,\mathbf{i}p_x + \mathbf{j}\,m) \qquad -RBG$$
$$(ikE - i\,\mathbf{j}p_y + \mathbf{j}\,m) \qquad +BRG$$
$$(ikE + i\,\mathbf{j}p_y + \mathbf{j}\,m) \qquad -GRB$$
$$(ikE + i\,\mathbf{k}p_z + \mathbf{j}\,m) \qquad +GBR$$
$$(ikE - i\,\mathbf{k}p_z + \mathbf{j}\,m) \qquad -BGR \qquad (35)$$

with the third and fourth notably changing the sign of the **p** component. The group structure required is clearly the one required by the conventional picture of 'coloured' quarks, that is an *SU*(3) structure, with eight generators and wavefunction

$$\psi \sim (BGR - BRG + GRB - GBR + RBG - RGB).$$

The 'colour' transitions in (34) could be seen as produced either by an exchange of the components of **p** between the individual quarks or baryon components, or as a relative switching of the component positions. That is, the colours could either move with the respective $p_x$, $p_y$, $p_z$ components, or switch with them. The two models contain exactly the same information, and also require a sign reversal in **p** as an additional consequence. If the **p** terms are regarded as operators, rather than as eigenvalues, they will be represented by the vector parts of the covariant derivatives required for an *SU*(3) local gauge transformation, the scalar part replacing the *E* term and incorporating the Coulomb part of the interaction.

The transition must be gauge invariant, because no direction is privileged, so the mediators must be massless, exactly as in the conventional picture, where the interaction is mediated by eight massless gluons. In this formulation, the gluons will be constructed from:

$$(\pm kE \mp ii\,\mathbf{i}p_x)(\mp kE \mp ii\,\mathbf{j}p_y) \quad (\pm kE \mp ii\,\mathbf{j}p_y)(\mp kE \mp ii\,\mathbf{i}p_x)$$
$$(\pm kE \mp ii\,\mathbf{j}p_y)(\mp kE \mp ii\,\mathbf{k}p_z) \quad (\pm kE \mp ii\,\mathbf{k}p_z)(\mp kE \mp ii\,\mathbf{j}p_y)$$
$$(\pm kE \mp ii\,\mathbf{i}p_z)(\mp kE \mp ii\,\mathbf{i}p_x) \quad (\pm kE \mp ii\,\mathbf{i}p_x)(\mp kE \mp ii\,\mathbf{i}p_z) \qquad (36)$$

and two combinations of

$$(\pm kE \mp ii\,\mathbf{i}p_x)(\mp kE \mp ii\,\mathbf{i}p_x) \quad (\pm kE \mp ii\,\mathbf{j}p_y)(\mp kE \mp ii\,\mathbf{j}p_y)$$
$$(\pm kE \mp ii\,\mathbf{k}p_z)(\mp kE \mp ii\,\mathbf{k}p_z) \qquad (37)$$

In addition to providing a quantum mechanical representation for baryon and gluon states, the structures derived in this section also suggest the existence of solutions to fundamental physical problems. The first is the mass-gap problem for baryons. In effect, why do baryons have nonzero mass and how can this mass be produced by the action of massless gluons? The structures in (34) and (35) clearly require the simultaneous existence of two states of helicity for the symmetry to remain unbroken, and this can only be possible if the baryon has nonzero mass. Further, this process is the signature of the Higgs mechanism, and so, contrary to much current supposition, the generation of the masses of baryons follows exactly the same process as that of all other fermions. However, this does not contradict the fact, established by much calculation using QCD, that the bulk of the mass of a baryon is due to the exchange of massless gluons, as the exchange of gluons structured as in (36) and (37) will necessarily lead to a sign change in the **p** operator, and hence of helicity, the exact mechanism which is responsible for the production of all known particle masses. In fact, the same will be true of all fermions involved in spin 1 boson exchange, and so all fermions must have nonzero masses.

The second problem is the specific nature and mechanism of the strong interaction between quarks. Again, the solution comes from the exact structure of the nilpotent operator. Here, we know, from (23), that there must be a Coulomb component or inverse linear potential ($\propto 1/r$), just to accommodate spherical symmetry. This has a known physical manifestation in the one-gluon exchange. But there is also at least one other component, which is responsible for quark confinement, for infrared slavery and for asymptotic freedom, and a linear potential ($\propto r$) has long been hypothesized and used in calculations. Here, we see that an exchange of **p** components at a constant rate, as in (34), would, in principle, require a constant rate of change of momentum, which is the signature of a linear potential.

In the nilpotent formalism, a differential operator incorporating Coulomb and linear potentials from a source with spherical symmetry (either the centre of a 3-quark system or one component of a quark-antiquark pairing) can be written in the form:

$$\left( \pm \boldsymbol{k}\left(E + \frac{A}{r} + Br\right) \mp i\left(\frac{\partial}{\partial r} + \frac{1}{r} \pm i\frac{j + \frac{1}{2}}{r}\right) + ijm \right). \tag{38}$$

If we can identify the phase factor to which this operator applies, to yield nilpotent solutions, it might be possible to show, for the first time on an analytic basis, that it is associated with a force which has characteristics identifiable with those of the strong interaction. By analogy with the pure Coulomb calculation, we might propose that the phase factor is of the form:

$$\phi = \exp(-ar - br^2) r^\gamma \sum_{\nu=0} a_\nu r^\nu ,$$

Applying the operator in (38) and the nilpotent condition, we obtain:

$$E^2 + 2AB + \frac{A^2}{r^2} + B^2 r^2 + \frac{2AE}{r} + 2BEr = m^2$$

$$-\left( a^2 + \frac{(\gamma + \nu + \ldots + 1)^2}{r^2} - \frac{(j + \frac{1}{2})^2}{r^2} + 4b^2 r^2 + 4abr - 4b(\gamma + \nu + \ldots + 1) - \frac{2a}{r}(\gamma + \nu + \ldots + 1) \right)$$

with the positive and negative $i(j + \frac{1}{2})$ terms cancelling out over the four solutions, as previously. Then, assuming a termination in the power series (as with the Coulomb solution), we can equate:

| coefficients of $r^2$ to give | $B^2 = -4b^2$ |
| coefficients of $r$ to give | $2BE = -4ab$ |
| coefficients of $1/r$ to give | $2AE = 2a(\gamma + v + 1)$ |

These equations immediately lead to:

$$b = \pm \frac{iB}{2}$$
$$a = \mp iE$$
$$\gamma + v + 1 = \mp iA.$$

The ground state case (where $v = 0$) then requires a phase factor of the form:

$$\phi = \exp(\pm iEr \mp iBr^2/2) r^{\mp iqA - 1}.$$

The imaginary exponential terms in $\phi$ can be seen as representing asymptotic freedom, the $\exp(\mp iEr)$ being typical for a free fermion. The complex $r^{\gamma - 1}$ term can be structured as a component phase, $\chi(r) = \exp(\pm iqA \ln(r))$, which varies less rapidly with $r$ than the rest of $\phi$. We can therefore write $\phi$ as

$$\phi = \frac{\exp(kr + \chi(r))}{r},$$

where
$$k = \pm iE \mp iBr/2.$$

The first term dominates at high energies, where $r$ is small, approximating to a free fermion solution, which can be interpreted as asymptotic freedom, while the second term, with its confining potential $Br$, dominates, at low energies, when $r$ is large, and this can be interpreted as infrared slavery. The Coulomb term, which is required to maintain spherical symmetry, is the component which defines the strong interaction phase, $\chi(r)$, and this can be related to the directional status of **p** in the state vector.

**8 Partitioning the vacuum**

In the nilpotent formalism, the characteristics of vacuum directly reflect those of matter, so we should expect to find that it has structure. If we take ($\pm i k E \pm i \mathbf{p} + j m$) and post-multiply it by the idempotent $k(\pm i k E \pm i \mathbf{p} + j m)$ any number of times, the only change is to introduce a scalar multiple, which can be normalized away.

($\pm i k E \pm i \mathbf{p} + j m$) $k(\pm i k E \pm i \mathbf{p} + j m)$ $k(\pm i k E \pm i \mathbf{p} + j m)$ … → ($\pm i k E \pm i \mathbf{p} + j m$) (39)

The same applies if we post-multiply by $i(\pm i k E \pm i \mathbf{p} + j m)$ or $j(\pm i k E \pm i \mathbf{p} + j m)$. The three idempotent terms have the mathematical properties of vacuum operators. However, another way of looking at (39) is to apply a time-reversal transformation to every even ($\pm i k E \pm i \mathbf{p} + j m$). Then we have ($\pm i k E \pm i \mathbf{p} + j m$) ($\mp\, i k E \pm i \mathbf{p} + j m$) ($\pm i k E \pm i \mathbf{p} + j m$) … $\rightarrow$ ($\pm i k E \pm i \mathbf{p} + j m$)  (40) with every even bracket becoming an antifermion, or combining with the original fermion state to become a spin 1 boson ($\pm i k E \pm i \mathbf{p} + j m$) ($\mp\, k E \pm i \mathbf{p} + j m$).

The same process can be applied using $i(\pm\, i k E \pm i \mathbf{p} + j m)$ and $j(\pm\, i k E \pm i \mathbf{p} + j m)$, and the result is that, from an initial fermion state, we generate either three vacuum reflections, via respective *T*, *P* and *C* transformations, which represent antifermion with the same spin, fermion with opposite spin, and antifermion with opposite spin, or combined particle-vacuum states which have the respective structures of spin 1 bosons, spin 0 bosons, or boson-like paired fermion (PF) combinations of the same kind as constitute Cooper pairs and the elements of Bose-Einstein condensates. Using just the lead terms of the nilpotents, we could represent these as:

($ikE$ + $i\mathbf{p}$ + $jm$) $k$ ($ikE$ + $i\mathbf{p}$ + $jm$) $k$ ($ikE$ + $i\mathbf{p}$ + $jm$) $k$ ($ikE$ + $i\mathbf{p}$ + $jm$) …   T  
($ikE$ + $i\mathbf{p}$ + $jm$) ($-ikE$ + $i\mathbf{p}$ + $jm$) ($ikE$ + $i\mathbf{p}$ + $jm$) ($-ikE$ + $i\mathbf{p}$ + $jm$) …   spin 1

($ikE$ + $i\mathbf{p}$ + $jm$) $j$ ($ikE$ + $i\mathbf{p}$ + $jm$) $j$ ($ikE$ + $i\mathbf{p}$ + $jm$) $j$ ($ikE$ + $i\mathbf{p}$ + $jm$) …   P  
($ikE$ + $i\mathbf{p}$ + $jm$) ($-ikE - i\mathbf{p}$ + $jm$) ($ikE$ + $i\mathbf{p}$ + $jm$) ($-ikE - i\mathbf{p}$ + $jm$) …   spin 0

($ikE$ + $i\mathbf{p}$ + $jm$) $i$ ($ikE$ + $i\mathbf{p}$ + $jm$) $i$ ($ikE$ + $i\mathbf{p}$ + $jm$) $i$ ($ikE$ + $i\mathbf{p}$ + $jm$) …   C  
($ikE$ + $i\mathbf{p}$ + $jm$) ($ikE - i\mathbf{p}$ + $jm$) ($ikE$ + $i\mathbf{p}$ + $jm$) ($ikE - i\mathbf{p}$ + $jm$) …   PF   (41)

So, we can repeatedly post-multiply a fermion operator by any of the discrete idempotent vacuum operators, creating an alternate series of antifermion and fermion vacuum states, or, equivalently, an alternate series of boson and fermion states without changing the character of the real particle state. Essentially a fermion produces a boson state by combining with its own vacuum image, and the two states form a supersymmetric partnership. Nilpotent operators are thus intrinsically supersymmetric, with supersymmetry operators typically of the form:

Boson to fermion:   $Q = \left(\pm ikE \pm i\mathbf{p} + jm\right)$  
Fermion to boson:   $Q^\dagger = \left(\mp ikE \pm i\mathbf{p} + jm\right)$

A fermion converts to a boson by multiplication by an antifermionic operator; a boson converts to a fermion by multiplication by a fermionic operator, and we could represent the first sequence in (41) by the supersymmetric

$$Q\, Q^\dagger\, Q\, Q^\dagger\, Q\, Q^\dagger\, Q\, Q^\dagger\, Q \ldots$$

We can choose to interpret this as the series of boson and fermion loops, of the same energy and momentum, required by the exact supersymmetry which would eliminate the need for renormalization, and remove the hierarchy problem altogether. Fermions and bosons (with the same values *E*, **p** and *m*) become their own supersymmetric partners through the creation of vacuum states, making the hypothesis of a set of real supersymmetric particles to solve the hierarchy problem entirely superfluous.

The identification of $i(ikE + i\mathbf{p} + jm)$, $k(ikE + i\mathbf{p} + jm)$ and $j(ikE + i\mathbf{p} + jm)$ as vacuum operators and $(ikE - i\mathbf{p} + jm)$, $(-ikE + i\mathbf{p} + jm)$ and $(-ikE - i\mathbf{p} + jm)$ as their respective vacuum 'reflections' at interfaces provided by *P*, *T* and *C* transformations suggests a new insight into the meaning of the Dirac 4-spinor. With the extra knowledge we have now gained, we can interpret the three terms other than the lead term *in the spinor* as the vacuum 'reflections' that are created with the particle. We can regard the existence of three vacuum operators as a result of a partitioning of the vacuum as a result of quantization and as a consequence of the 3-part structure observed in the nilpotent fermionic state, while the *zitterbewegung* can be taken as an indication that the vacuum is active in defining the fermionic state.

Taken together, the four components of the spinor cancel exactly, especially when represented as operators using discrete calculus, as in (19). The four components can be represented as creation operators for

| | |
|---|---|
| fermion spin up | $(ikE + i\mathbf{p} + jm)$ |
| fermion spin down | $(ikE - i\mathbf{p} + jm)$ |
| antifermion spin down | $(-ikE + i\mathbf{p} + jm)$ |
| antifermion spin up | $(-ikE - i\mathbf{p} + jm)$ |

or annihilation operators for

| | |
|---|---|
| antifermion spin down | $(ikE + i\mathbf{p} + jm)$ |
| antifermion spin up | $(ikE - i\mathbf{p} + jm)$ |
| fermion spin up | $(-ikE + i\mathbf{p} + jm)$ |
| fermion spin down | $(-ikE - i\mathbf{p} + jm)$ |

They could equally well be regarded as two operators for creation and two for annihilation, for example:

| | |
|---|---|
| fermion spin up creation | $(ikE + i\mathbf{p} + jm)$ |
| fermion spin down creation | $(ikE - i\mathbf{p} + jm)$ |
| fermion spin up annihilation | $(-ikE + i\mathbf{p} + jm)$ |
| fermion spin down annihilation | $(-ikE - i\mathbf{p} + jm)$ |

Either way, the cancellation is exact, both physically, and algebraically (when we use the discrete operators which leave out the passive mass component). It is interesting that the cancellation requires *four* components, rather than two, for, while the transitions:

$$(ikE + i\mathbf{p} + jm) \to (ikE - i\mathbf{p} + jm)$$
and
$$(ikE + i\mathbf{p} + jm) \to (-ikE + i\mathbf{p} + jm)$$

can occur through spin 1 boson and spin 0 paired fermion exchange, and the active space and time components, there is no process in nature for the *direct* transition:

$$(i k E + i \mathbf{p} + j m) \rightarrow (-i k E - i \mathbf{p} + j m)$$

with no active component as agent. In this context, it might be worth noting that the spin 0 fermion-fermion state

$$(i k E + i \mathbf{p} + j m)\,(i k E - i \mathbf{p} + j m)$$

is such as would be required in a pure weak transition from $-ikE$ to $+ikE$, or its inverse.

Because the formation of the spin 0 state necessarily requires intrinsically massive components, even in those cases where it assumes nonzero effective mass through a Fermi velocity less than $c$, time reversal symmetry (the one applicable to the transition) must be broken in the weak formation or decay of such states. The most likely opportunity of observing such a process might be in one of the physical manifestations of the nonzero Berry phase, say the quantum Hall effect, in some special type of condensed matter such as graphene. Here, the conduction electrons have zero effective mass and a Hamiltonian that can be written in the form $\pm v_F \mathbf{i}(\mathbf{i} p_x + \mathbf{j} p_y)$, where $v_F$ is the Fermi velocity. We can imagine creating a boson-like state with single-valued spin by the quantum Hall effect, Aharonov-Bohm effect, or Bose-Einstein condensation, and then observing, perhaps through a change in the Fermi velocity during its decay, the violation of both $P$ and $CP = T$ symmetries.

## 9 Supersymmetry and renormalization

If exact supersymmetry is a consequence of the nilpotent formalism and its representation of vacuum, then a free fermion in vacuum should produce its own loop cancellations and its energy should acquire a finite value without renormalization. Free fermion plus boson loops should cancel, and there should be no hierarchy problem. We can examine this possibility by performing a basic perturbation calculation for first order coupling in QED, and showing that it leads to zero in the case of a free fermion. Suppose we have a fermion acted on by the electromagnetic potentials $\phi$, $\mathbf{A}$. Then, using only the lead terms of the spinors for simplicity,

$$\left(-\mathbf{k}\frac{\partial}{\partial t} - i\mathbf{i}\nabla + \mathbf{j}m\right)\psi = -e\left(+i\mathbf{k}\phi + i\mathbf{i}\mathbf{A}\right)\psi$$

We now apply a perturbation expansion to $\psi$, so that

$$\psi = \psi_0 + \psi_1 + \psi_2 + \dots,$$

with

$$\psi_0 = (i\mathbf{k}E + i\mathbf{i}\mathbf{p} + \mathbf{j}m)\,e^{-i(Et - \mathbf{p}\cdot\mathbf{r})}$$

as the solution of the unperturbed equation:

$$\left(-k\frac{\partial}{\partial t}-ii\nabla+jm\right)\psi=0,$$

which represents zeroth-order coupling, or a free fermion of momentum **p**.

Using the perturbation expansion, we can write

$$\left(-k\frac{\partial}{\partial t}-ii\nabla+jm\right)(\psi_0+\psi_1+\psi_2+...)=-e(k\phi+ii\mathbf{A})(\psi_0+\psi_1+\psi_2+...),$$

from which we can extract the first-order coupling as

$$\left(-k\frac{\partial}{\partial t}-ii\nabla+jm\right)\psi_1=-e(k\phi+ii\mathbf{A})\psi_0.$$

If we expand ($k\phi$ + $ii\mathbf{A}$) as a Fourier series, and sum over momentum **k**, we obtain

$$(k\phi+ii\mathbf{A})=\Sigma\,(k\phi(\mathbf{k})+ii\mathbf{A}(\mathbf{k}))\,e^{i\mathbf{k}\cdot\mathbf{r}},$$

so that

$$\left(-k\frac{\partial}{\partial t}-ii\nabla+jm\right)\psi_1=-e\sum(k\phi(\mathbf{k})+ii\mathbf{A}(\mathbf{k}))e^{i\mathbf{k}\cdot\mathbf{r}}\psi_0$$

$$=-e\sum(k\phi(\mathbf{k})+ii\mathbf{A}(\mathbf{k}))e^{i\mathbf{k}\cdot\mathbf{r}}(ikE+i\mathbf{p}+jm)e^{-i(Et-\mathbf{p}\cdot\mathbf{r})}$$

$$=-e\sum(k\phi(\mathbf{k})+ii\mathbf{A}(\mathbf{k}))(ikE+i\mathbf{p}+jm)e^{-i(Et-(\mathbf{p}+\mathbf{k})\cdot\mathbf{r})}$$

Suppose we now expand $\psi_1$ as

$$\psi_1=\sum v_1(E,\mathbf{p}+\mathbf{k})e^{-i(Et-(\mathbf{p}+\mathbf{k})\cdot\mathbf{r})}$$

Then

$$\sum\left(-k\frac{\partial}{\partial t}-ii\nabla+jm\right)v_1(E,\mathbf{p}+\mathbf{k})e^{-i(Et-(\mathbf{p}+\mathbf{k})\cdot\mathbf{r})}$$

$$=-e\sum(k\phi(\mathbf{k})+ii\mathbf{A}(\mathbf{k}))(ikE+i\mathbf{p}+jm)e^{-i(Et-(\mathbf{p}+\mathbf{k})\cdot\mathbf{r})}$$

and

$$\sum(ikE+i(\mathbf{p}+\mathbf{k})+jm)v_1(E,\mathbf{p}+\mathbf{k})e^{-i(Et-(\mathbf{p}+\mathbf{k})\cdot\mathbf{r})}$$

$$=-e\sum(k\phi(\mathbf{k})+ii\mathbf{A}(\mathbf{k}))(ikE+i\mathbf{p}+jm)e^{-i(Et-(\mathbf{p}+\mathbf{k})\cdot\mathbf{r})}$$

and, equating individual terms,

$$(ikE+i(\mathbf{p}+\mathbf{k})+jm)\,v_1(E,\mathbf{p}+\mathbf{k})=-e\,(k\phi(\mathbf{k})+ii\mathbf{A}(\mathbf{k}))\,(ikE+i\mathbf{p}+jm).$$

We can write this in the form

$$v_1(E,\mathbf{p}+\mathbf{k})=-e[ikE+i(\mathbf{p}+\mathbf{k})+ijm]^{-1}\,(k\phi(\mathbf{k})+ii\,\mathbf{A}(\mathbf{k}))(ikE+i\mathbf{p}+jm)$$

which means that

$$\psi_1 = -e\sum \left[ikE + i(\mathbf{p}+\mathbf{k}) + jm\right]^{-1}\left(k\phi(\mathbf{k}) + ii\mathbf{A}(\mathbf{k})\right)\left(ikE + i\mathbf{p} + jm\right)e^{-i(Et-(\mathbf{p}+\mathbf{k})\cdot\mathbf{r})}$$

which is the wavefunction for first-order coupling, with a fermion absorbing or emitting a photon of momentum **k**.

However, if we observe the process in the rest frame of the fermion and eliminate any *external* source of potential, then **k** = 0, and ($k\phi$ + $ii\mathbf{A}$) reduces to the static value, $k\phi$, with $\phi$ as a self-potential. In this case, $\psi_1$ becomes

$$\psi_1 = -e\left[ikE + i\mathbf{p} + jm\right]^{-1}(k\phi)(ikE + i\mathbf{p} + jm)e^{-i(Et-\mathbf{p}\cdot\mathbf{r})} ,$$

as the summation is no longer strictly required for a single order of the pure self-interaction. Since we can also write this as

$$\psi_1 = -e(ikE + i\mathbf{p} + jm)(ikE + i\mathbf{p} + jm)(k\phi)e^{-i(Et-\mathbf{p}\cdot\mathbf{r})} , \qquad (42)$$

we see that $\psi_1 = 0$, for any fixed value of $\phi$. Clearly, this will also apply to higher orders of self-interaction In other words, a *non-interacting* nilpotent fermion requires no renormalization as a result of its self-energy. In addition, the process could also be adapted for interacting particles, subject to external potentials, for here we can imagine redefining the *E* and **p** operators to incorporate external potentials to make them 'internal', while simultaneously changing the structure of the phase factor to accommodate this. The change of phase factor would, of course, require a corresponding change in the amplitude, which could be taken as redetermining the value of the coupling constant, *e*, as required. A possibly more rigorous calculation would use the cancellation method to pair off fermion and boson loops with exact supersymmetry. Ultimately, however, (42) shows that it is the structure of (*ikE* + *i***p** + *jm*) as a nilpotent which seemingly eliminates the infinite self-interaction terms in the perturbation expansion at the same time as showing that they are merely an expression of the nature of the nilpotent vacuum as a reflection of the exactly supersymmetric nature of the original particle state.

**10 Propagators**

If the nilpotent formalism can, in principle, remove a divergence with a seemingly physical origin, then it should also readily remove a divergence which is generally considered to be a purely mathematical artefact. Here, we are considering the 'pole' or singularity that appears in the propagator in the conventional Feynman formalism. This is a classic sign of the action of vacuum, and it is generally taken as the point of 'switchover' between fermion and antifermion states. The nilpotent formalism, however, removes the pole automatically because of its direct inclusion of vacuum states. Conventional theory assumes that a fermion propagator takes the form

$$S_F(p) = \frac{1}{\not{p} - m} = \frac{\not{p} + m}{p^2 - m^2},$$

where $\not{p}$ represents $\gamma^\mu \partial_\mu$, or its eigenvalue, and that there is a singularity or 'pole' ($p_0$) where $p^2 - m^2 = 0$, the 'pole' being the origin of positron states. On either side of the pole there are positive energy states moving forwards in time, and negative energy states moving backwards in time, the terms ($\not{p} + m$) and ($-\not{p} + m$) being used to project out, respectively, the positive and negative energy states. The normal solution is to add an infinitesimal term $i\varepsilon$ to $p^2 - m^2$, so that $iS_F(p)$ becomes

$$iS_F(p) = \frac{i(\not{p}+m)}{p^2 - m^2 + i\varepsilon} = \frac{(\not{p}+m)}{2p_0}\left(\frac{1}{p_0 - \sqrt{p^2 + m^2} + i\varepsilon} + \frac{1}{p_0 + \sqrt{p^2 + m^2} - i\varepsilon}\right)$$

and take a contour integral over the complex variable to give the solution

$$S_F(x - x') = \int d^3p \frac{1}{(2\pi)^3} \frac{m}{2E}\left[-i\theta(t-t')\sum_{r=1}^{2}\Psi(x)\overline{\Psi}(x') + i\theta(t'-t)\sum_{r=3}^{4}\Psi(x)\overline{\Psi}(x')\right]$$

with summations over the up and down spin states.

This mathematical subterfuge is unnecessary in the nilpotent formalism because the denominator of the propagator term is always a nonzero scalar. We write

$$S_F(p) = \frac{1}{(\pm ikE \pm i\mathbf{p} + jm)},$$

and choose our usual interpretation of the reciprocal of a nilpotent to give:

$$\frac{1}{(\pm ikE \pm i\mathbf{p} + jm)} = \frac{(\pm ikE \mp i\mathbf{p} - jm)}{(\pm ikE \pm i\mathbf{p} + jm)(\pm ikE \mp i\mathbf{p} - jm)} = \frac{(\pm ikE \mp i\mathbf{p} - jm)}{4(E^2 + p^2 + m^2)},$$

which is finite at all values. The integral is now simply

$$S_F(x - x') = \int d^3p \frac{1}{(2\pi)^3} \frac{m}{2E} \theta(t - t') \Psi(x) \overline{\Psi}(x'),$$

in which $\Psi(x)$ is the usual

$$\Psi(x) = (\pm ikE \pm i\mathbf{p} + jm) \exp(ipx),$$

with the phase factor written as a 4-vector, and the adjoint term becomes

$$\overline{\Psi}(x') = (\pm ikE \mp i\mathbf{p} - jm)(ik) \exp(-ipx').$$

Since the nilpotent formalism comes as a complete package with a single phase term, automatic second quantization, and negative time states matched with reverse time states, there is no averaging over spin states or separation of positive and negative energy states on opposite sides of a pole.

The fermion propagator can also be used to define boson propagators. In conventional theory, we derive the boson propagator (43) directly from the Klein-Gordon equation, while recognizing that its mathematical form depends on the choice of gauge:

$$\Delta_F(x - x') = \frac{\slashed{p} + m}{p^2 - m^2}. \tag{43}$$

This is because the Klein-Gordon operator

$$\left(\gamma^0 \frac{\partial}{\partial t} + \boldsymbol{\gamma}.\nabla + im\right)\left(\gamma^0 \frac{\partial}{\partial t} + \boldsymbol{\gamma}.\nabla - im\right) = \left(\frac{\partial^2}{\partial t^2} - \nabla^2 + m^2\right)$$

is the only scalar product which can emerge from a differential operator defined as in (1). The Klein-Gordon equation, however, is not specific to boson states or an identifier of them. It merely defines a universal zero condition which is true for all states, whether bosonic or fermionic. And, the propagator in (43) does not correspond to any known bosonic state. Instead, we have *three* boson propagators.

Spin 1: $\quad \Delta_F(x - x') = \dfrac{1}{(\pm ikE \pm i\mathbf{p} + jm)(\mp ikE \pm i\mathbf{p} + jm)},$

Spin 0: $\quad \Delta_F(x - x') = \dfrac{1}{(\pm ikE \pm i\mathbf{p} + jm)(\mp ikE \mp i\mathbf{p} + jm)},$

Paired Fermion: $\quad \Delta_F(x - x') = \dfrac{1}{(\pm ikE \pm i\mathbf{p} + jm)(\pm ikE \mp i\mathbf{p} + jm)}.$

Where the spin 1 bosons are massless (as in QED), we will have expressions like:

$$\Delta_F(x - x') = \frac{1}{(\pm ikE \pm i\mathbf{p})(\mp ikE \pm i\mathbf{p})}. \tag{44}$$

Clearly, the relationship of the fermion and boson propagators is of the form

$$S_F(x - x') = (i\gamma^\mu \partial_\mu + m)\, \Delta_F(x - x'),$$

or, in our notation,

$$S_F(x - x') = (\pm ikE \pm i\mathbf{p} + jm)\, \Delta_F(x - x').$$

which is exactly the same relationship as is defined between fermion and boson in the nilpotent formalism.

Now, using

$$iS_F(p) = \frac{1}{2p_0}\left(\frac{1}{p_0 - \sqrt{p^2 + m^2} + i\varepsilon} + \frac{1}{p_0 + \sqrt{p^2 + m^2} - i\varepsilon}\right),$$

which is the same as the conventional fermion propagator up to a factor ($\not{p} + m$), we can perform a contour integral which is similar to that for the fermion to produce

$$i\Delta_F (x - x') = \int d^3p \frac{1}{(2\pi)^3} \frac{1}{2\omega} \theta(t - t') \phi(x)\phi^*(x').$$

Here, $\omega$ takes the place of $E / m$, while $\phi(x)$ and $\phi(x')$ are now scalar wavefunctions. However, in our notation, they will be scalar products of ($\pm i k E \pm i\mathbf{p} + jm$) exp ($ipx$) and ($\mp i k E \pm i\mathbf{p} + jm$) exp ($ipx'$) and $\phi(x)\phi^*(x')$ reduces to a product of a scalar term, which can be removed by normalization, and exp $ip(x - x')$.

In off-mass-shell conditions, where $E^2 \neq p^2 + m^2$, poles in the propagator are a mathematical, rather than physical, problem, and removed by the use of $i\varepsilon$ and the contour integral, which is *ad hoc* but effective. However, in the specific case of massless bosons, conventional theory cannot prevent 'infrared' divergencies appearing in (43) when such bosons are emitted from an initial or final stage which is on the mass shell. Such divergencies, however, do not occur where there is no pole, as in (44). The definition of the boson propagator as (44), rather than (43), not only shows that one of the principal divergences in quantum electrodynamics is, as the procedure used to remove it would suggest, merely an artefact of the mathematical structure we have imposed, and not of a fundamentally physical nature, but also suggests that the formalism which removes it is a more exact representation of the fundamental physics. Ultimately, this is because it allows an exact representation of the vacuum simultaneously with the fermionic state.

**11 Weak interactions**

So far, we have been able to show that two fundamental interactions are *intrinsic to the definition of the fermionic state*, and not something external imposed upon it. The Coulomb interaction, as we see from (24), is the direct product of spherical symmetry being applied at the same time as nilpotency. Since it is purely an expression of the magnitude of a scalar phase, all the terms in the nilpotent contribute, but only one, the passive (scalar) mass term, contributes to nothing else. An interaction with this precise property may therefore be defined, and it is the one we define as the *electric* interaction. At the same time, the strong interaction, with its characteristic linear potential, can be represented as we have seen, by the vector properties of the **p** term. However, yet another interaction seems to be required by the *spinor* structure of the nilpotent operator, and the associated phenomenon of *zitterbewegung*. While the co-existence of two spin states is, in some sense, real, and is accounted for by the presence of mass, the co-existence of two energy states is only meaningful in the context of the simultaneous existence of fermion and vacuum. While the transitions between the two energy states may be virtual, in this sense, the *zitterbewegung* would seem to require the production of an intermediate bosonic state at a vertex where one fermionic state is annihilated and another is created to replace by it. This behaviour is, of course, characteristic of the weak interaction, and, in this sense, we can say that the weak interaction, like the electric and strong interactions, is built into the structure of the nilpotent operator.

The weak interaction is clearly related to the nature of the pseudoscalar $iE$ operator, whose sign uniquely determines the helicity of a weakly interacting particle, or more specifically its weakly interacting component. It also has a unique feature, in that its fermionic source cannot be separated from its vacuum partner. A fermion or antifermion cannot be created or annihilated, even with an antifermionic or fermionic partner, unless its vacuum is simultaneously annihilated or created. In this sense, the weak source has a manifestly dipolar nature, whose immediate manifestation is the fermion's ½-integral spin. This, then, leads to the question of whether we can derive an analytic expression from the nilpotent operator which will explain the special characteristics of this force. To answer this, it will be convenient to answer a more general question: what nilpotent solutions are available for an operator including a Coulomb potential together with any other potential which is a function of distance from a point source with spherical symmetry, other than the linear potential characteristic of the strong interaction?

We will assume that the nilpotent operator takes a form such as

$$\left( k\left( E - \frac{A}{r} - Cr^n \right) + i\left( \frac{\partial}{\partial r} + \frac{1}{r} \pm i\frac{j+\frac{1}{2}}{r} \right) + ijm \right). \tag{45}$$

where $n$ is an integer greater than 1 or less than $-1$, and, as usual, look for a phase factor which will make the amplitude nilpotent. Again, we will work from the basis of the Coulomb solution, with the additional information that polynomial potential terms which are multiples of $r^n$ require the incorporation into the exponential of terms which are multiples of $r^{n+1}$. So, extending our work on the Coulomb solution, we may suppose that the phase factor is of the form:

$$\phi = \exp(-ar - br^{n+1})r^\gamma \sum_{\nu=0} a_\nu r^\nu$$

Applying the operator and squaring to zero, with a termination in the series, we obtain

$$4\left( E - \frac{A}{r} - Cr^n \right)^2 = -2\left( -a + (n+1)br^n + \frac{\gamma}{r} + \frac{\nu}{r} + \frac{1}{r} + i\frac{j+\frac{1}{2}}{r} \right)^2$$
$$-2\left( -a + (n+1)br^n + \frac{\gamma}{r} + \frac{\nu}{r} + \frac{1}{r} - i\frac{j+\frac{1}{2}}{r} \right)^2$$

Equating constant terms, we find
$$a = \sqrt{m^2 - E^2} \tag{46}$$
Equating terms in $r^{2n}$, with $\nu = 0$:
$$C^2 = -(n+1)^2 b^2$$
$$b = \pm \frac{iC}{(n+1)}.$$

Equating coefficients of *r*, where *v* = 0:

$$AC = -(n+1)\,b\,(1+\gamma),$$
$$(1+\gamma) = \pm iA.$$

Equating coefficients of $1/r^2$ and coefficients of $1/r$, for a power series terminating in $v = n'$, we obtain

$$A^2 = -(1+\gamma+n')^2 + (j+\tfrac{1}{2})^2 \qquad (47)$$

and

$$-EA = a(1+\gamma+n'). \qquad (48)$$

Combining (46), (47) and (48) produces:

$$\left(\frac{m^2 - E^2}{E^2}\right)(1+\gamma+n')^2 = -(1+\gamma+n')^2 + (j+\tfrac{1}{2})^2$$

$$E = -\frac{m}{j+\tfrac{1}{2}}(\pm iA + n'). \qquad (49)$$

Equation (49) has the form of a harmonic oscillator, with evenly spaced energy levels deriving from integral values of *n'*. If we make the additional assumption that *A*, the phase term required for spherical symmetry, has some connection with the random directionality of the fermion spin, we might assign to it a half-unit value ($\pm \tfrac{1}{2} i$), or ($\pm \tfrac{1}{2} i\hbar c$), using explicit values for the constants, and obtain the complete formula for the fermionic simple harmonic oscillator:

$$E = -\frac{m}{j+\tfrac{1}{2}}(\tfrac{1}{2} + n'). \qquad (50)$$

The dimensions of *A* are those of charge (*q*) squared or interaction energy × range, and an *A* numerically equal to $\pm \tfrac{1}{2} \hbar c$ would be exactly that required by the uncertainty principle, allowing the value of the range of an interaction mediated by the Z boson to be calculated as $\hbar / 2M_Z c = 2.166 \times 10^{-18}$ m, as observed. The $\tfrac{1}{2} \hbar c$ term is also significant in the expressions for zero-point energy and *zitterbewegung*, which connect with both spin and the uncertainty principle. Interpreting the *zitterbewegung* as a dipolar switching between fermion and vacuum antifermion states, we can describe this in terms of a weak dipole moment $(\hbar c / 2)^{3/2} / M_Z c^2$, of magnitude $8.965 \times 10^{-18}$ *e* m. Because of the specific appearance of the $\tfrac{1}{2} \hbar c$ term for spin (*s*) in $\mu = gqs / 2m$, an identical expression can additionally be used to define a weak magnetic moment, of order $4.64 \times 10^{-5}$ × the magnetic moment of the electron. The possible appearance of an imaginary factor *i* in *A* is interesting in relation to the requirement of a complex potential or vacuum for *CP* violation in the pure weak interaction.

Whatever assumptions we make about *A*, equation (49) demonstrates that the additional potential of the form $Cr^n$, where *n* is an integer greater than 1 or less than – 1, has the effect of creating a harmonic oscillator solution for the nilpotent operator, irrespective of the value of *n*, and, in fact, we can show that any polynomial sum of potentials of this form will produce the same result. Such potentials emerge from any system in which there is complexity, aggregation, or a multiplicity of sources, even if the individual sources have Coulomb or linear potentials. In the case of a dipolar weak sources, there will be a minimum extra term of the form $Cr^{-3}$, and so we can say that (49) provides the correct characteristics for the weak interaction from the kind of potential that weak sources must necessarily produce. In addition, because this solution is exclusive for distance related potentials of the form $Cr^n$, except where *r* = 1 or −1, we have also, in effect, shown that a fermion interaction specified in relation to a spherically symmetric point source has only three physical manifestations, and that these are the ones associated with the electric (or other pure Coulomb), strong and weak interactions.

**12 Vacuum partitions and sources**

In the previous section, we have seen that the nilpotent operator (± *ikE* ± *i***p** + *jm*), with its three components, *iE*, **p** and *m*, separated by three quaternion units ***k***, ***i*** and ***j***, is a source of interactions with the characteristics we describe as weak, strong and electric, *through its own structure*. We have also identified the origins of these interactions in the structures of the components. All the interactions contribute to the mass and dual spin state, and the magnitudes of all three terms, but only the weak interaction identifies the energy state or the sign attached to the pseudoscalar component *iE*, and the active direction of handedness, and only the strong interaction can respond to the vector or directional aspect of **p**. The electric interaction is distinguishable by *only* responding to the magnitude scale in the same way as *m*. It may be significant here that the Kaluza-Klein theory, which effectively introduces a fifth dimension with a *U*(1) symmetry, which has a similar role to that played by mass in the nilpotent structure, is actually two separate theories with the respective aims of explaining the origins of mass and electric charge.

In principle, the association of the three quaternion units with objects with different mathematical properties, which are themselves connected with the physical parameters time, space and mass, suggests that the vacuum partitions responding to the operators ***k***(± *ikE* ± *i***p** + *jm*), ***i***(± *ikE* ± *i***p** + *jm*) and ***j***(± *ikE* ± *i***p** + *jm*) are to some extent those which create the physical effects connected with the respective weak, strong and electric interactions, and that the units ***k***, ***i*** and ***j***, to this extent, act as the sources. The object of these 'sources' is then to create the vacuum partitions that lead to the physical manifestations of the interactions in individual nilpotent structures. Of course, this picture takes no account of gravity, but it is yet to be established that gravity is a local force, like the others, or that it is determined by discrete sources. It is just as likely that it is a vacuum effect, 'disguised' as a local force by the inertial effect which it inevitably produces. The universal ubiquity, negative energy and positive norm of the coupling constant suggest fundamental differences with respect to the other interactions. In this case, the vacuum for gravity could be $\pm 1(\pm ikE \pm i\mathbf{p} + j m)$, equivalent to the first term in the nilpotent. Clearly, there are significant aspects of gravity yet to be explained, in particular, dark matter and dark energy, and it is not yet established whether a quantum theory of gravity is actually possible. Further exploration of the vacuum can be expected to lead to a greater understanding of this difficult matter.

In addition to gravity, string theory has already been mentioned as an area where vacuum has a significantly active role. A perfect string theory is believed to be one in which self-duality in phase space determines vacuum selection. Interestingly, the nilpotent $(\pm ikE \pm i\mathbf{p} + jm)$ is an object which has all these required characteristics. It also has a '10-dimensional' structure in that the 5 'dimensions' of $E$, $\mathbf{p}$, $m$ are paralleled by the 5 source terms, $k$, $3 \times i$ terms, and $j$; and 6 of these (all but $E$ and $\mathbf{p}$) are conserved quantities, and, in that sense, 'compacted'. It may be that an exploration of vacuum in these terms might also produce significant results in particle structures.

**13 Conclusion**

Vacuum is identified in this paper as the state that remains when a fermion is created out of absolutely nothing. It is an existence condition, like conservation of energy, which must apply irrespective of our knowledge of how it is constructed, and it can be defined with exact mathematical precision. Like other existence conditions, the mathematical definition of the nilpotent vacuum provides a constraint upon the physical possibilities. This allows us to construct a version of quantum mechanics, which is more compact and powerful than any other formalism, as can be seen from work in earlier publications [8-18]. It is already second quantized and intrinsically supersymmetric, and eliminates many currently perceived problems such as the infrared divergence, the mass gap and the hierarchy problem. Incorporating vacuum directly into the mathematics at the same time as the fermion effectively doubles the information available and halves the information to be discovered. It also gives us a way of seeing aspects of the Standard Model as automatic consequences of the nilpotent formalism. None of this actually gives the exact structure of the vacuum, in the sense of constructing the 'rest of the universe' that needs to exist to make a fermion in a particular state actually possible. However, it does suggest that the explanation of some things that are currently mysterious, such as dark matter, dark energy, and even gravity itself, might respond, at some future date, to considerations based on the physical requirements that are needed to maintain the nilpotent vacuum existence condition.